\documentclass[manuscript]{aastex63}
\shortauthors{bora et al.}

\usepackage{amsmath,amssymb}

\usepackage{hyperref}

\usepackage{graphicx}
\begin{document}
\title{Evolution of three-dimensional coherent structures in Hall magnetohydrodynamics}
\author{K. Bora} 
\affil{Udaipur Solar Observatory, Physical Research Laboratory, Dewali, Bari Road, Udaipur-313001, India}
\affil{Discipline of Physics, Indian Institute of Technology, Gandhinagar-382355, India}
\author{R. Bhattacharyya}
\affil{Udaipur Solar Observatory, Physical Research Laboratory, Dewali, Bari Road, Udaipur-313001, India}

\author{P. K. Smolarkiewicz}
\affil{National Center for Atmospheric Research, Boulder, Colorado, USA}

\accepted{on November, 6 2020}

\begin{abstract}
The work extends the computational model EULAG-MHD to include Hall
magnetohydrodynamics (HMHD)---important to explore
physical systems undergoing fast magnetic reconnection at the
order of the ion inertial length scale.  Examples include
solar transients along with reconnections in magnetosphere, magnetotail  and
 laboratory plasmas. The paper documents the results  
of two distinct sets of implicit large-eddy
simulations in the presence and absence of the Hall forcing term, 
initiated with an unidirectional sinusoidal magnetic field. 
The HMHD simulation while benchmarking the code also emphasizes the
complexity of three dimensional (3D) evolution over its two dimensional
(2D) counterpart. The  magnetic reconnections onset significantly earlier in HMHD. 
Importantly,  the magnetic field generated by the Hall term breaks any inherent symmetry, ultimately making the evolution 3D. The resulting 3D reconnections develop magnetic
flux ropes and magnetic flux tubes. Projected on the reconnection plane,
the ropes and tubes appear as magnetic islands, which later break into
secondary islands, and finally coalesce to generate an X-type neutral
point. These findings are in agreement with the theory and contemporary
simulations of HMHD, and thus verify our extension of the EULAG-MHD model.
The second set explores the influence of the Hall forcing on generation and 
ascend of a magnetic flux rope 
from sheared magnetic arcades---a novel scenario instructive in understanding 
the coronal transients. The rope
evolves through intermediate complex structures, ultimately
breaking locally because of reconnections. Interestingly,
the breakage occurs earlier in the presence of the Hall term,
signifying faster dynamics leading to magnetic topology favorable
for reconnections.
\end{abstract}



\section{Introduction}
\label{intro}

Magnetofluids characterized by large Lundquist numbers $S=LV_A/\eta$
($L\equiv$~length scale of the the magnetic field ${\bf{B}}$ variability,
$V_A \equiv$~Alfv\'en speed, and $\eta \equiv$~magnetic diffusivity)
satisfy Alfv\'en's flux-freezing theorem on magnetic field lines (MFLs)
being tied to fluid parcels \citep{Alfven}.  The astrophysical plasmas
are of particular interest, because their inherently large $L$ implies
large $S$ and the flux freezing. For example, the solar corona with
global $L\approx 10^6~{\rm m}$, $V_A\approx 10^6~{\rm m}~{\rm s}^{-1}$,
$B\approx 10~{\rm G}$ and $\eta\approx 1~{\rm m}^2~{\rm s}^{-1}$
(calculated using Spitzer resistivity) has a $S\approx 10^{12}$
\citep{Aschwanden}.  Nevertheless the coronal plasma also exhibits
diffusive behavior. For example, the solar transients---such as flares,
coronal mass ejections, and coronal jets---are all manifestations of
magnetic reconnections (MRs) that in turn form a diffusive phenomenon,
where the magnetic energy gets converted into heat and kinetic energy of
plasma flow, accompanied with a rearrangement of MFLs \citep{Arnab}.
The onset of MRs is due to the generation of small scales in consequence of large scale dynamics, 
eventually 
leading to locally reduced characteristic length scale of the magnetic field variability and thereby resulting 
in intermittently diffusive plasmas. 
 The small scales can owe their origin to the 
presence of magnetic nulls; i.e., locations where the 
${\bf{B}}=0$ \citep{MHDpriest,Ss2020}. Alternatively, and more relevant to this
paper, this can be due to the presence of current sheets (CSs); i.e., the
ribbons of intense current, across which ${\bf{B}}$ has a sharp gradient
\citep{ParkerECS,DKRB}. Spontaneous development of CSs is predicted by
the Parker's magnetostatic theorem that implies the inevitability of CSs
in an equilibrium magnetofluid with perfect electrical conductivity. This
inevitability in turn is due to the general failure of a smooth magnetic
field to simultaneously preserve the local force balance and the global
magnetic topology \citep{ParkerECS}.

To elucidate this seminal theorem, we follow the arguments in
\citet{DKRB}. Let us perceive a series of contiguous fluid parcels with
their frozen-in MFLs. We further assume the fluid to be incompressible.
The magnetic field is everywhere continuous. If the two ends of the
series are pushed toward each other, because of incompressibility the
interstitial parcels will be squeezed out. Terminally, the end parcels of
the series will approach each other and their MFLs being non-parallel (in
general)---will create a CS. Notably, for an ideal magnetofluid having
infinite electrical conductivity the creation of the CS is the terminal
state, but in the presence of a small magnetic diffusivity the MFLs will
ultimately reconnect. The reconnected MFLs are subsequently pushed away
from the reconnection region by the outflow, and as they come out of
the CS, the MFLs once again get frozen to the fluid. Afterward, these
frozen MFLs can push another flux system and repeat the whole process
of reconnection. This switch between the large and the small scales with
their inherent coupling is fundamentally interesting and has the potential
to drive a myriad of MR driven phenomena observed in the solar atmosphere.

An example of this scenario is numerically demonstrated by \citet{Sanjay2016}
using MHD simulations. Their simulations describe the activation of a
magnetic flux rope (MFR) by an interplay between the two aforementioned
scales. Notably the large scale is relatively independent of the
particular system under consideration but identifying the small or
the diffusion scale depends on the specific physical system involved.
For instance, observations suggests the average MR time for solar flares
to be $10^2-10^3$ s; the impulsive rise time \citep{PF200}. Presuming
the relation $L\equiv\sqrt{\tau_d\eta}$ holding well and the magnetic
diffusion time scale $\tau_d\approx 10^3$ s, the $L$ that initiates
the MR turns out to be $\approx 32$~m.  The local Lundquist number
then gets reduced to $S\approx 3.2\times 10^7$.  As a consequence,
an ion inertial scale $\delta_i\approx 2.25$ m in the solar
corona \citep{PF200} suggests that the order of the dissipation term,
$1/S\approx 10^{-7}$, is much smaller than the order of the Hall term,
$\delta_i/ L\approx 10^{-2}$, in an appropriate dimensionless induction
equation\footnote{Hereafter, a constant electron number density is assumed.}
\citep{Westerberg}
\begin{equation}
\label{inducresist}
\frac{{\partial\bf{B}}}{\partial t} = 
\nabla\times \left({\bf{v}}\times{\bf{B}}\right)
-\frac{1}{S}\nabla\times{\bf{J}}
-\frac{\delta_i}{L}\nabla\times\left({\bf{J}}\times{\bf{B}}\right)~,
\end{equation}
where ${\bf{J}}(=\nabla\times{\bf{B}})$ and ${\bf{v}}$ are the volume current density and 
the plasma flow velocity, respectively. The disparity in the magnitude orders of the dissipation and the Hall
forcing suggests that if the dissipation is important, then so is the
Hall magnetohydrodynamics (HMHD). It also indicates that HMHD can be 
crucial for coronal transients---MRs being their underlying reason. 
By the same token, HMHD is also important in other systems like Earth's
magnetosphere, typically at the magnetopause and the magnetotail where
CSs exist \citep{Mozer2002}.

In the HMHD, the ion and electron motions decouple \citep{Sonnerup},
and the MFLs are frozen in the electron fluid instead of the ion fluid.
Additionally, straightforward mathematical manipulations show that the
Hall term in the induction equation does not affect the dissipation
rate of magnetic energy and magnetic helicity \citep{PF200}. Given the
unique properties of HMHD, we expect it to reveal subtle changes in
MFLs evolution familiar from the standard MHD \citep{Sanjay2016, Ss2019, Ss2020}, and refine the dynamics leading to magnetic reconnections. To
further explore HMHD specifically contextual to the solar physics,
we have extended the EULAG-MHD model \citep{PiotrJCP,Piotrscience}
by including the Hall forcing and document here the results pertaining
to two distinct sets of large-eddy simulations. The first set focuses
on benchmarking the numerical model by verifying the HMHD physics in
three spatial dimensions. The simulations highlight the complexity in
reconnection-assisted formation of coherent magnetic structures, hitherto
less explored in the contemporary research. The second set, simulates, for
the first time to our knowledge, the formation and evolution of a magnetic
flux rope (MFR) under the influence of the Hall forcing---initiated from
sheared bipolar magnetic loops relevant to the solar corona.

Related to our simulations, \citet{Mozer2002} using 2D geometry have shown
that the Hall forcing causes the electrons residing on the reconnection
plane containing the MFLs to flow into and out of the reconnection
region, generating the in-plane current.  The in-plane current, in turn,
develops a magnetic field having component out of the reconnection
plane. This out-of-plane magnetic field, or the Hall magnetic field,
has quadrupole structure. The asymmetric propagation of the reconnection
plane, because of a ``reconnection wave'' has been shown by \citet{Huba}
in a simulation of 3D MR in the Hall limit, where curved out-of-plane
MFLs along with the density gradient play a crucial role.

The remainder of the paper is organized as follows. Section~\ref{ivp}
outlines the numerical model. Section~\ref{results} benchmarks the
code using 3D simulations and, subsequently presents a novel numerical
experiment that compares activation of a solar-like MFR under standard MHD
and HMHD formulations. Section~\ref{conclusion} summarizes the key
findings of this work.

\section{The numerical model}
\label{ivp}

A numerical simulation consistent with physics of solar corona must accurately 
preserve the flux-freezing by minimizing 
numerical dissipation and dispersion errors away from the reconnection 
regions characterized by steep gradients of the magnetic field \citep{RBCLOW}.
 Such minimization is a signature of a class of
inherently nonlinear high-resolution transport methods that preserve
field extrema along flow trajectories, while ensuring higher-order
accuracy away from steep gradients in advected fields. Consequently,
we incorporate the Hall forcing in the established high-resolution
EULAG-MHD model \citep{PiotrJCP, Piotrscience}, a specialized version
of the general-purpose hydrodynamic model EULAG predominantly used in
atmospheric and climate research \citep{Prusa}. Central to the EULAG is
the spatio-temporally second-order-accurate nonoscillatory forward-in-time
(NFT) advection scheme MPDATA, a.k.a Multidimensional Positive Definite
Advection Transport Algorithm, \citep{Piotrsingle}. A feature unique
to MPDATA and important in our calculations is its widely-documented
dissipative property that mimics the action of explicit subgrid-scale
turbulence models, wherever the concerned advective field is
under-resolved---the property referred to as implicit large eddy simulations
(ILES) \citep{Grinstein2007}. The resulting MRs remove the under-resolved
scales and restore the flux-freezing. These MRs being intermittent and
local, successfully mimic physical MRs. The ILES property of MPDATA have proven
instrumental in a series of advanced numerical studies across a range of
scales and physical scenarios, including studies related to the coronal
heating along with data-constrained simulations of solar flares and
coronal jets \citep{RBCLOW, DKRB, SKRB, SK2017, avijeet2017, avijeet2018,
Ss2019, Ss2020}. The simulations reported in this paper also benefit from
ILES property of MPDATA.

Here, the numerically integrated HMHD equations assume a perfectly conducting, 
incompressible magnetofluid. Using a conservative flux-form and dyadic 
notation, they are compactly written (assuming cgs units) as
\begin{eqnarray}
\label{momtransf}
\frac{\partial{\bf v}}{\partial t} + \nabla\cdot{\bf v}{\bf v}&=&
 -\nabla \phi + \frac{1}{4\pi\rho_0}\nabla\cdot{\bf B}{\bf B} + \mu_0\nabla^2{\bf v}~,\\
\label{induc}
\frac{\partial{\bf B}}{\partial t} + \nabla\cdot{\bf v}{\bf B}&=&
\nabla\cdot{\bf B}{\bf v}-\frac{d_0}{4\pi}\left(\nabla\times\nabla\cdot{\bf B}{\bf B}\right)
 -\nabla \phi^*~,\\
\label{incompv}
\nabla\cdot {\bf v}&=& 0~, \\
\label{incompb}
\nabla\cdot {\bf B}&=& 0~,
\end{eqnarray}
where $\rho_0$ and $\mu_0$ denote, respectively, constant density and
kinematic viscosity, $\phi=\left(p+\textbf{B}^2/8\pi\right)/\rho_0$ is the
density normalized total pressure, and $d_0=\sqrt{4\pi}\delta_i/\rho_0$;
the $-\nabla \phi^*$ term on the right-hand-side (rhs) of~(\ref{induc}) 
will be explained shortly.

To highlight the numerics of EULAG-MHD and its extension to HMHD,
the prognostic PDEs (\ref{momtransf}) and (\ref{induc}) are further
symbolized as a single equation
\begin{equation}
\label{eul}
\frac{\partial{\bf \Psi}}{\partial t}+\nabla \cdot{\bf v\Psi}={\bf R}{\bf\Psi}~,
\end{equation}
where ${\bf \Psi}=\{v_x, v_y, v_z, B_x, B_y, B_z\}^T$ is the vector of
prognosed variables, and ${\bf R}({\bf \Psi})$ is the vector of their
associated rhs forcings. The principal second-order-accurate NFT Eulerian
algorithm for (\ref{eul}) can be written compactly as
\begin{equation}
\label{nft}
{\bf \Psi}_{\bf i}^n={\mathcal A}_{\bf i}\left({\bf{{\Psi}}}^{n-1}+\delta_ht\,
{\bf R}^{n-1}, {\bf v}^{n-1/2}\right) +\delta_ht\,{\bf R}({\bf\Psi})\vert^n_{\bf i}
~\equiv~ \widehat{\bf\Psi}_{\bf i}+\delta_ht\,{\bf R}({\bf\Psi})\vert^n_{\bf i}~,
\end{equation}
where $n$, ${\bf i}$ refer to $(t^n,{\bf x}_{\bf i})$ locations
on a regular collocated grid, $\delta t$ marks the time step with
$\delta_ht=0.5\delta t$, and ${\cal{A}}$ denotes the MPDATA advection
operator, solely dependent on the preceding, $t^{n-1}=t^n-\delta t$,
values of ${\bf\Psi}$ as well as a first-order estimate of the solenoidal
velocity at $t^n-\delta_ht$ extrapolated from the earlier values.

The principal algorithm (\ref{nft}) is implicit for all prognosed
variables and diagnosed potentials $\phi$ and $\phi^*$ that enter
(\ref{momtransf}) and (\ref{induc}), respectively. While $\phi$
is a physical variable, $\phi^*$ is a numerical facilitator enabling
restoration of (\ref{incompb})---viz. divergence cleaning---eventually
polluted with truncation errors. Because of its nonlinearity,
the rhs ${\bf{R}}$ of (\ref{nft}) is viewed as a combination
of a linear term ${\bf{L}}{\bf\Psi}$ (with ${\bf L}$ denoting a
known linear operator), a nonlinear term ${\bf N}({\bf\Psi})$, and
the potential term $-\nabla{\bf{\Phi}}$ with ${\bf{\Phi}} \equiv
(\phi,~\phi,~\phi,~\phi^*,~\phi^*,~\phi^*)^T$.  The resulting form of
(\ref{nft}) is realized iteratively with the nonlinear part of the rhs
forcing lagged behind,
\begin{equation}
\label{outtmpl}
{\mathbf{\Psi}}^{n,\nu}_{\bf i}= {\widehat {\mathbf{\Psi}}}_{\bf i}
 + \delta_ht\ {\mathbf{L}}{\mathbf{\Psi}}\big|^{n,\nu}_{\bf i}
 + \delta_ht\ {\mathbf{N}}({\mathbf{\Psi}})\big|^{n,\nu-1}_{\bf i}
-\delta_ht\ \nabla{\mathbf{\Phi}}\big|^{n,\nu}_{\bf i}~,
\end{equation}
where $\nu=1,..,m$ numbers the iterations. The algorithm in
(\ref{outtmpl}) is still implicit with respect to
${\bf{\Psi}}^{n,\nu}_{\bf i}$ and ${\bf{\Phi}}^{n,\nu}_{\bf i}$,
yet straightforward algebraic manipulations lead to the closed-form
expression 
\begin{equation}
\label{outsolv}
{\mathbf{\Psi}}^{n,\nu}_{\bf i}= \left[{\mathbf{I}}-\delta_ht
{\mathbf{L}}\right]^{-1} \left({\widehat {\widehat {\mathbf{\Psi}}}}-\delta_ht
{\nabla}{\mathbf{\Phi}}^{n,\nu}\right)\Big|_{\bf i}~,
\end{equation}
where ${\widehat{\widehat{\mathbf{\Psi}}}}\equiv{\widehat{\mathbf{\Psi}}}
+\delta_ht {\mathbf{N}}{\mathbf{\Psi}}\big|^{n,\nu-1}$ denotes the modified 
explicit element of the solution. Taking the divergences of the first and the 
second three components of (\ref{outsolv}), produces two elliptic Poisson 
problems, for $\phi$  and $\phi^*$, respectively, as implied by (\ref{incompv}) 
and (\ref{incompb}).

The iterative formulation of (\ref{nft}) in (\ref{outtmpl}), outlines
the concept of the EULAG-MHD discrete integrals. The actual iterative
implementation of (\ref{nft}), detailed in \citep{PiotrJCP}, proceeds
in a sequence of steps such that the most current update of a dependent
variable is used in the ongoing step, wherever possible. Furthermore,
to enhance the efficacy of the scheme, judicious linearization of ${\bf
N}({\mathbf \Psi})$ is employed, together with a blend of evolutionary
and conservative forms of the induction and Lorentz forces.  Each outer
iteration has two distinct blocks. The focus of the first,``hydrodynamic''
block is on integrating the momentum equation, where the magnetic field
enters the Lorentz force and is viewed as supplementary.  This block
ends with the final update of the velocity via the the solution of the
elliptic problem for pressure. The second, ``magnetic'' block uses the
current updates of the velocities to integrate the induction equation.
It ends with the final update of the magnetic field via the solution of
the elliptic problem for the divergence cleaning. Incorporating the Hall
forcing into the EULAG-MHD model follows the principles of the outlined
standard MHD integrator. Because the Hall term enters (\ref{induc}) as
the curl of the Lorentz force, it can be judiciously updated and combined
with the standard induction forcing, whenever the Lorentz force and/or
the magnetic field are updated. In the current implementation it enters
the explicit (lagged) counterpart of the induction force, and is updated
after the inversion of the implicit evolutionary form of the induction
equation in the ``magnetic'' block; cf. section 3.2 in \citep{PiotrJCP}
for details.

\section{Results}
\label{results}

\subsection{Benchmarking the 3D HMHD solver}
To benchmark the HMHD solver, the initial field is selected as 
\begin{eqnarray}
\label{sinb1}
& & B_{x}=0~, \\
& & B_{y}=0~, \\
\label{sinb2}
& & B_{z}=2.5 \sin(x)~,
\label{sin3}
\end{eqnarray}
with $x, y, z \in[-2\pi,2\pi]$, respectively, in each direction of a 
3D Cartesian domain. This selection has two merits: first, the 
magnetic field reverses at $x$=0; and second, the Lorentz force 
\begin{eqnarray}
\label{lorentz1}
& & ({\bf{J}}\times{\bf{B}})_x =-6.25 \cos (x) \sin (x) \hspace{0.1cm}~, \\
& & ({\bf{J}}\times{\bf{B}})_y = 0~, \\
& & ({\bf{J}}\times{\bf{B}})_z = 0~,
\label{lorentz3}
\end{eqnarray}
generates a converging flow that onsets MRs. Being different
from the traditional initial conditions involving the Harris current
sheet or the GEM challenge \citep{BIRN}, this selection shows the
Hall effects are independent of particular initial conditions. We have 
explored simulations using the traditional initial conditions 
(not shown) and the outcomes are similar.

The equations (\ref{momtransf})-(\ref{incompb}) are integrated numerically,
as described in the preceding section, for $d_0=0,~2$. The
latter selection of $d_0$ optimizes the computation time and
a tractable development of magnetic structures for the employed 
spatio-temporal resolution. The corresponding $\delta_i=0.56$ is slightly higher than the spatial stepsize $\delta {\bf{x}}\approx 0.40$ set for the simulation.
Consequently, the Hall forcing kicks in  
near the dissipation scale, thereby directly 
affecting the overall dynamics only in  vicinities of the MR regions.
With the large scale $L=4 \pi$ of the magnetic field variability, 
the resulting $\delta_i/L\approx 0.04$ is on the order of solar coronal value.
The simulations are then expected to 
capture dynamics of  the HMHD and the intermittently diffusive regions of 
corona-like plasmas, thus shedding light on 
the evolution of neighboring frozen-in MFLs.
Moreover, $1/S < \delta_i/L$ as discussed in
Introduction.  The physical
domain is resolved with $32\times 32\times 32$ grid. A coarse resolution is 
selected for an earlier onset of MRs and to expedite the overall evolution.
The kinematic viscosity and mass density are set to $\nu=0.005$ and
$\rho_0=1$, respectively. All three boundaries are kept open. The initial
magnetic field is given by the equations (\ref{sinb1}-\ref{sin3}) and
the fluid is evolved from an initially static state having pressure $p=0$. 
The simulation parameters are listed in Table 1. 

\begin{figure*}[h]
\includegraphics[width=1\textwidth]{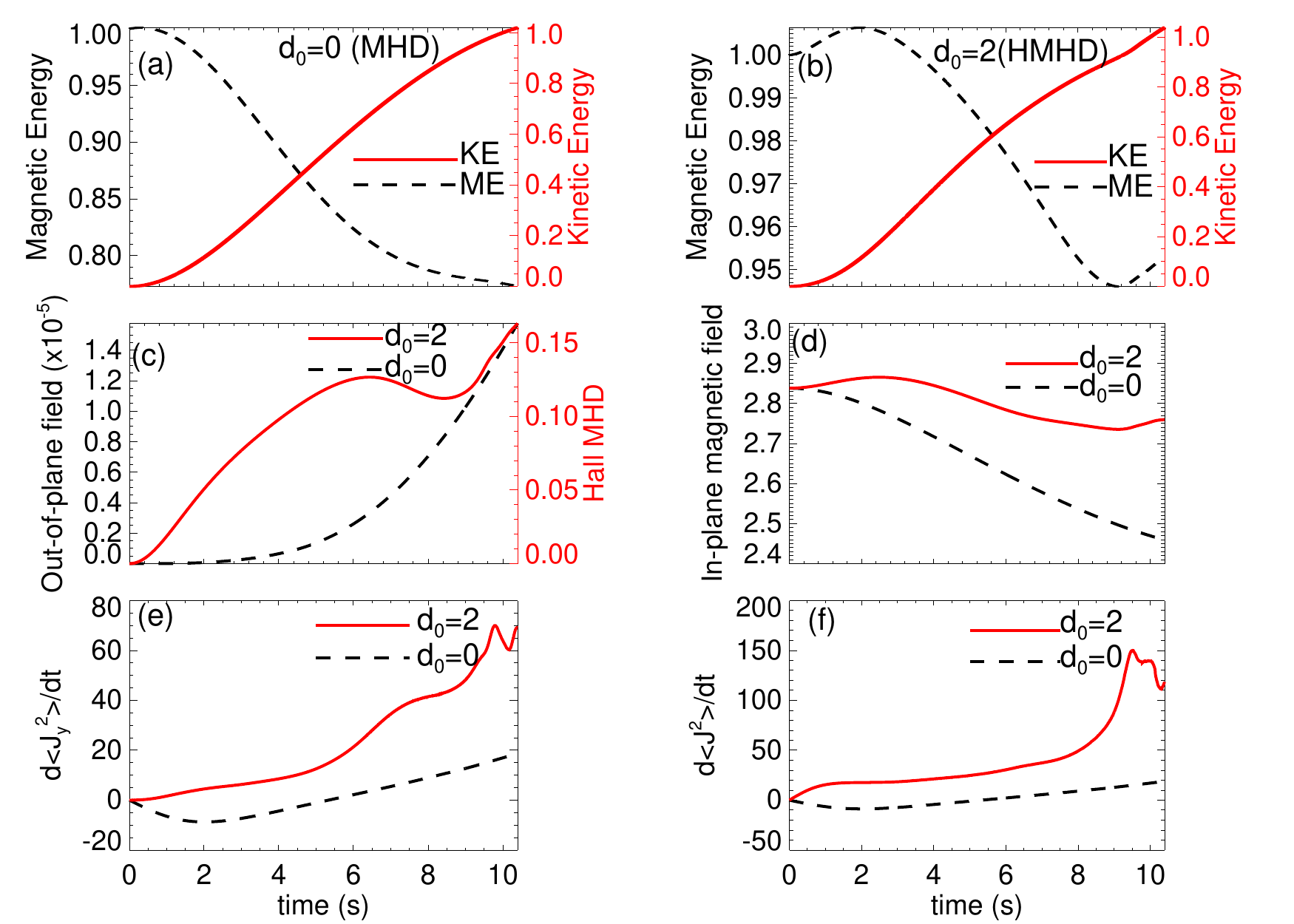}
\caption{Panels (a) and (b) show the evolution of the magnetic energy (black dashed curve) and kinetic energy (red solid curve) for $d_0=0$ (MHD) and $d_0=2$ (HMHD) respectively. Panel (c) shows the evolution of out-of-plane magnetic field for $d_0=0$ (MHD) with black dashed curve and $d_0=2$ (HMHD) with red solid curve respectively. Also in panels (a) to (c), the scales for the solid and the dashed curves are spaced at right and left respectively.  Panels (d) to (f) represent in-plane magnetic field, amplitudes of the rate of change of 
out-of-plane and total current densities for $d_0=0$ (black dashed curve) and $d_0=2$ (red solid curve) respectively. The variables in panels (a) and (b) are normalized with the initial total energies. All the variables are averaged over the computational domain. Important are the generation of the out-of-plane magnetic field along with sharp changes
in time derivatives of the out-of-plane and total volume current densities in HMHD simulations.
}\label{Sineoverall}
\end{figure*}

\begin{deluxetable*}{cccccccc}[h]
\tablenum{1}
\tablecaption{List of parameters for simulation with sinusoidal initial field\label{tab:tabi}}
\tablewidth{0pt}
\tablehead{
\colhead{$\rho_0$} & \colhead{$d_0$} & \colhead{$\delta_i$} & \colhead{L} 
& \colhead{$\frac{\delta_i}{L}$} & \colhead{Simulation Box Size} & \colhead{Resolution} &\colhead{$\nu$}
}
\startdata
1.0 & 2.0 & 0.56 & $4\pi$ & 0.04 & $(4\pi)^3$ & $(32)^3$ & 0.005
\enddata
\end{deluxetable*}
The overall evolution is depicted in different panels of Figure \ref{Sineoverall}. The initial Lorentz force, given by equations (\ref{lorentz1}-\ref{lorentz3}), pushes segments of the fluid on either sides of the field reversal layers---toward
each other. Consequently, magnetic energy gets converted into kinetic energy of the plasma flow: panels
(a) and (b).  Panels
(c) and (d) show history of  grid-averaged magnitude of the the out-of-plane (along $y$) and in-plane ($xz$ plane) magnetic fields. 
Notably, for $d_0=0$ (MHD) the out-of-plane field is negligibly small compared to its value for $d_0=2$ (HMHD). Such 
generation of the out-of-plane magnetic field is inherent to HMHD and is in 
conformity with the result of another simulation \citep{MaBhattacharjee}. The panels (e) and (f) illustrate
the variation of the rate of change of out-of-plane current density and total volume current density. 
Importantly, 
in contrast to the $d_0=0$ curve, the rate of change of volume current density shows an early bump at $(\approx 7.5$ s) and a 
well 
defined peak ($t\approx 9.75$ s) for $d_0=2$. Such peaks in the current density are expected in the impulsive phase
of solar flares, and they manifest MRs in the presence of the Hall term \citep{BhattacharjeeReview}.

\begin{figure*}[h]
\includegraphics[width=1\textwidth, height=0.8\textwidth]{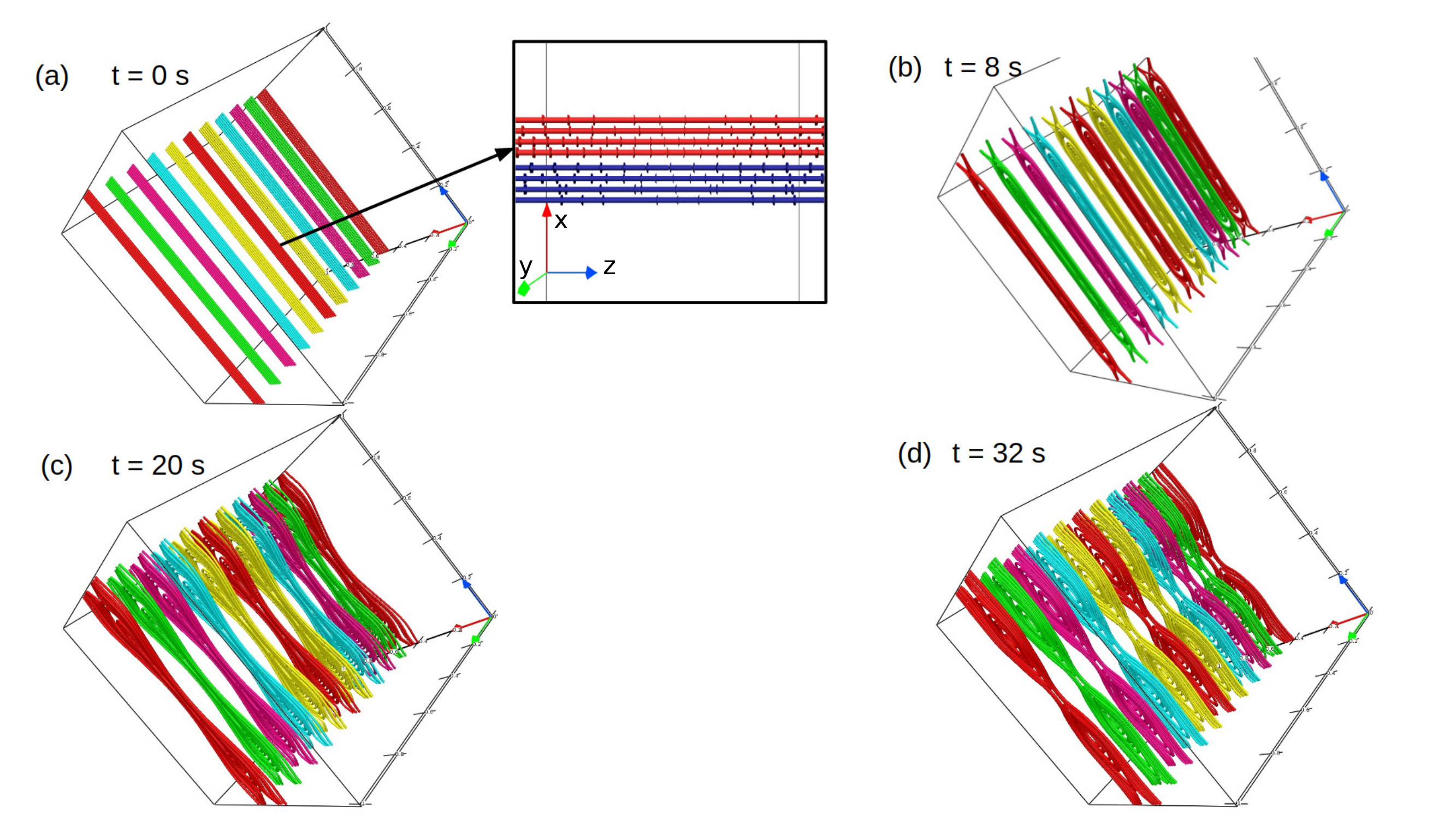}
\caption{Snapshots of preselected MFLs for the $d_0=0$ (MHD) simulation, plotted on equidistant y-constant planes. In all figures (this and hereafter), the 
red, green and blue arrows represent the x, y and z-axis respectively. The inset in panel (a) highlights the polarity reversal of the initial magnetic field lines. The plots illustrate 
the formation of a primary flux tube (panel (b)) made by stacking of the depicted MFLs.
Notably symmetry is preserved throughout evolution. (An animation of this figure is available.)
}\label{d_00concatenated}
\end{figure*}

\begin{figure*}[h]
\includegraphics[width=1\textwidth, height=0.8\textwidth]{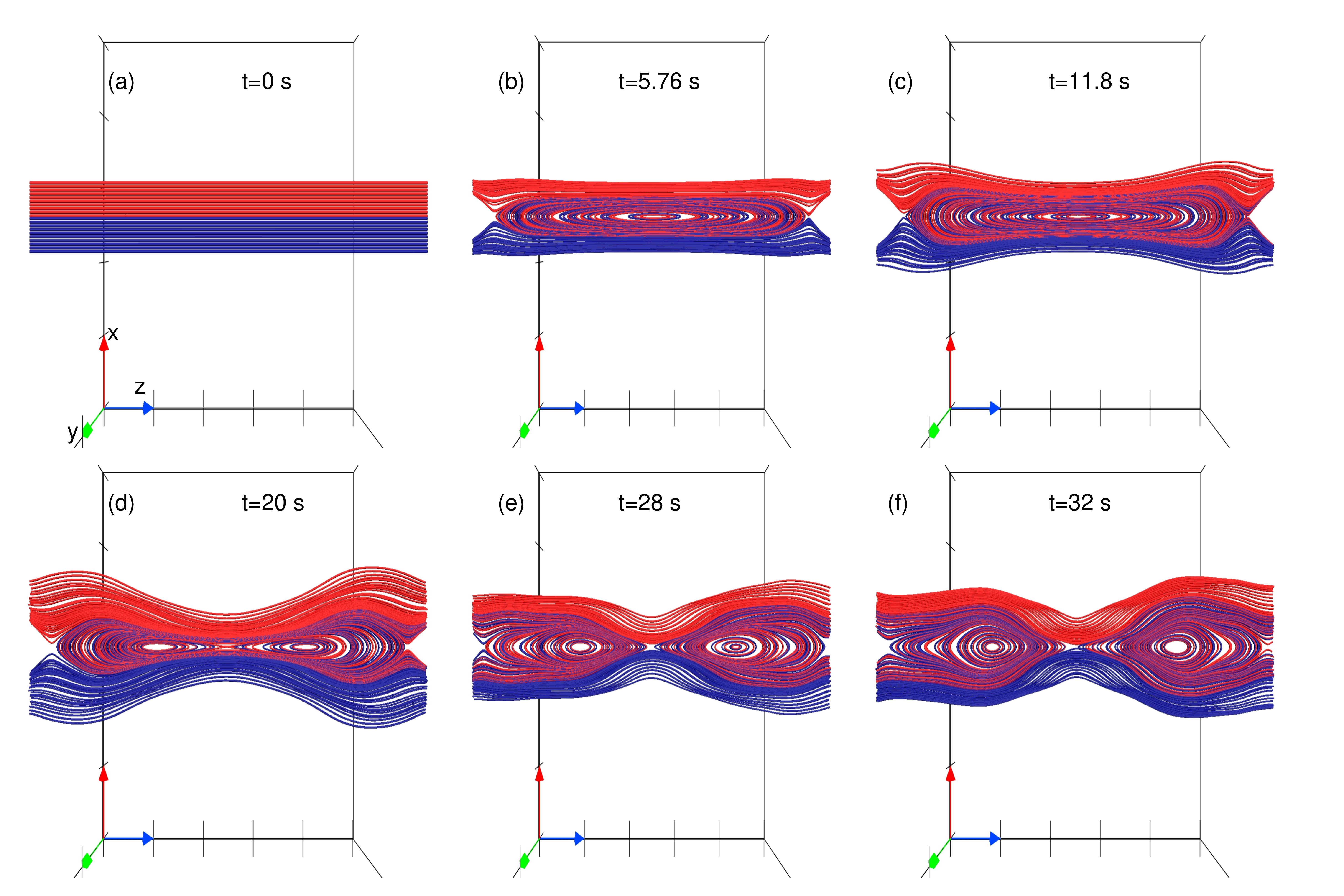}
\caption{Projection of MFLs depicted in Figure \ref{d_00concatenated}
on a $y$ constant plane during their evolution. Notable
is the formation of a primary magnetic island having a single O-type neutral point. 
Subsequently, the primary island breaks into two secondary islands which are
separated by an X type neutral point.}\label{2Dsind00proj}
\end{figure*}

Figure \ref{d_00concatenated}  plots MFLs tangential to pre-selected planes 
during different instances of the evolution for $d_0=0$. The panel (a) plots
 the initial MFLs for referencing. The initial Lorentz force pushes anti-parallel
MFLs (depicted in the inset) toward each other.  
Subsequently, X-type neutral points develop near $z=\pm 2\pi$. The consequent MRs generate a complete magnetic island which maintains its identity for a long time. Such islands, stacked on each other along the $y$, 
generate an extended magnetic flux tube (MFT) at the center, which in its generality is a magnetic 
flux surface. Further evolution breaks the MFT such that the cross section of the broken 
tube yields two magnetic islands. The point of contact between the two tear-drop shaped MFLs 
generates an X-type neutral point. Notably, within the computational time, 
no field is generated along the $y$ direction and the corresponding symmetry is 
exactly preserved. In Figure \ref{2Dsind00proj} we provide the 2D projection of the MFLs on 
the $y=0.5$ plane, for later comparison with similar projection for the $d_0=2$ case.

\begin{figure*}[h]
\includegraphics[width=1\textwidth]{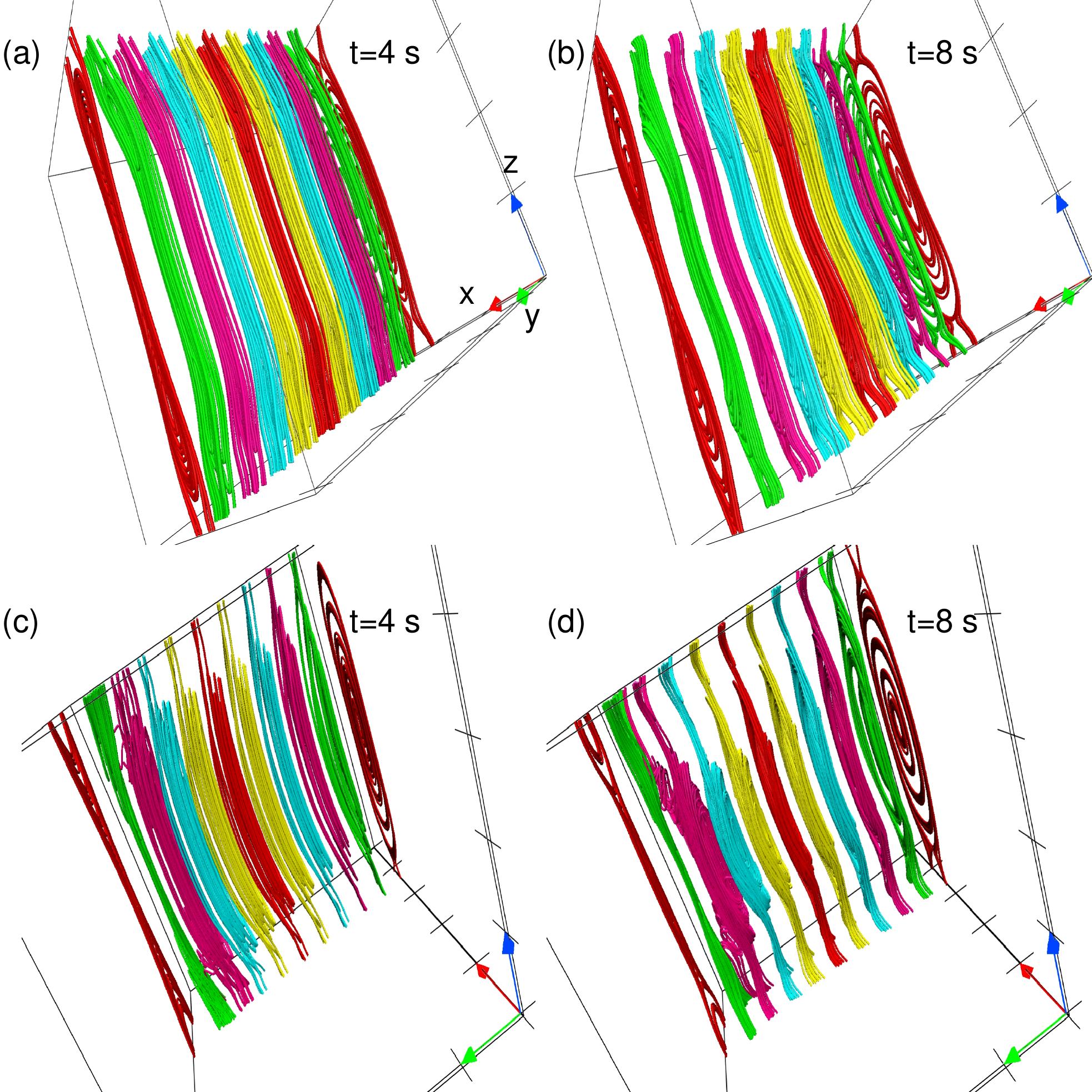}
\caption{MFL evolution for $d_0=2$ (HMHD) case, from two vantage points. The field lines 
are plotted on planes centered at $x=0.5$ and $x=0.74435$ and equidistant along $y$.
Important is the symmetry breaking, cf. MFLs at $y=-2\pi$ and $y=2\pi$ of
the panel (b) and (d). The  out-of-plane magnetic field is
generated throughout the domain. (A combined animation showing MFL evolution for the top and the bottom panels is available.) }\label{twovp}
\end{figure*}

\begin{figure*}[h]
\includegraphics[width=1\textwidth]{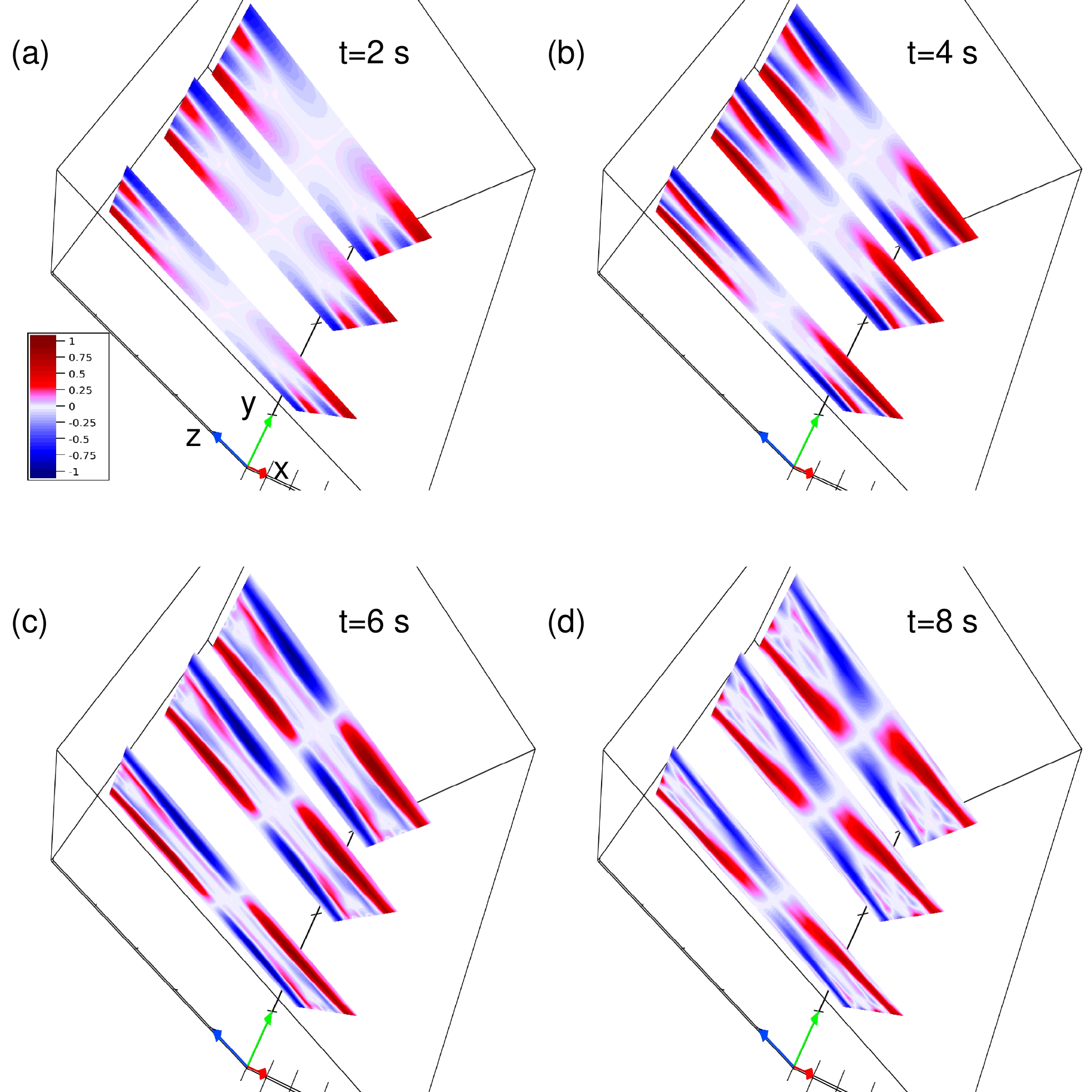}
\caption{Contour plots of $B_y(x,z)$ (out-of-plane component) on y-constant planes for $d_0=2$ (HMHD), with time. The plots confirm the quadrupolar nature of the out-of-plane component of the magnetic field.}\label{BQP}
\end{figure*}

The panels (a) to (b) and (c) to (d) 
of Figure \ref{twovp} show MFL evolution for $d_0=2$ from two different
vantage points. The MFLs are plotted on different $y$-constant planes centered at $x=0.5$ and $x=0.74435$. The planes are not connected
by any field lines at $t=0$. Importantly out-of-plane magnetic field  is generated with time in both sets of 
MFLs (at $x=0.5$ and $x=0.74435$), which connects two adjacent planes (cf. panel (b) of Figures \ref{d_00concatenated} and \ref{twovp}) and breaks the $y$-symmetry that was preserved in the $d_0=0$ case --- asymmetry in reconnection planes. 
Consequently the evolved ${\bf{B}}$ is three-dimensional. Also, the out-of-plane component ($B_y$) has a
quadrupole structure, shown in Figure \ref{BQP}, which is in congruence with observations and 
models \citep{Mozer2002}. 

\begin{figure*}[h]
\includegraphics[width=1\textwidth]{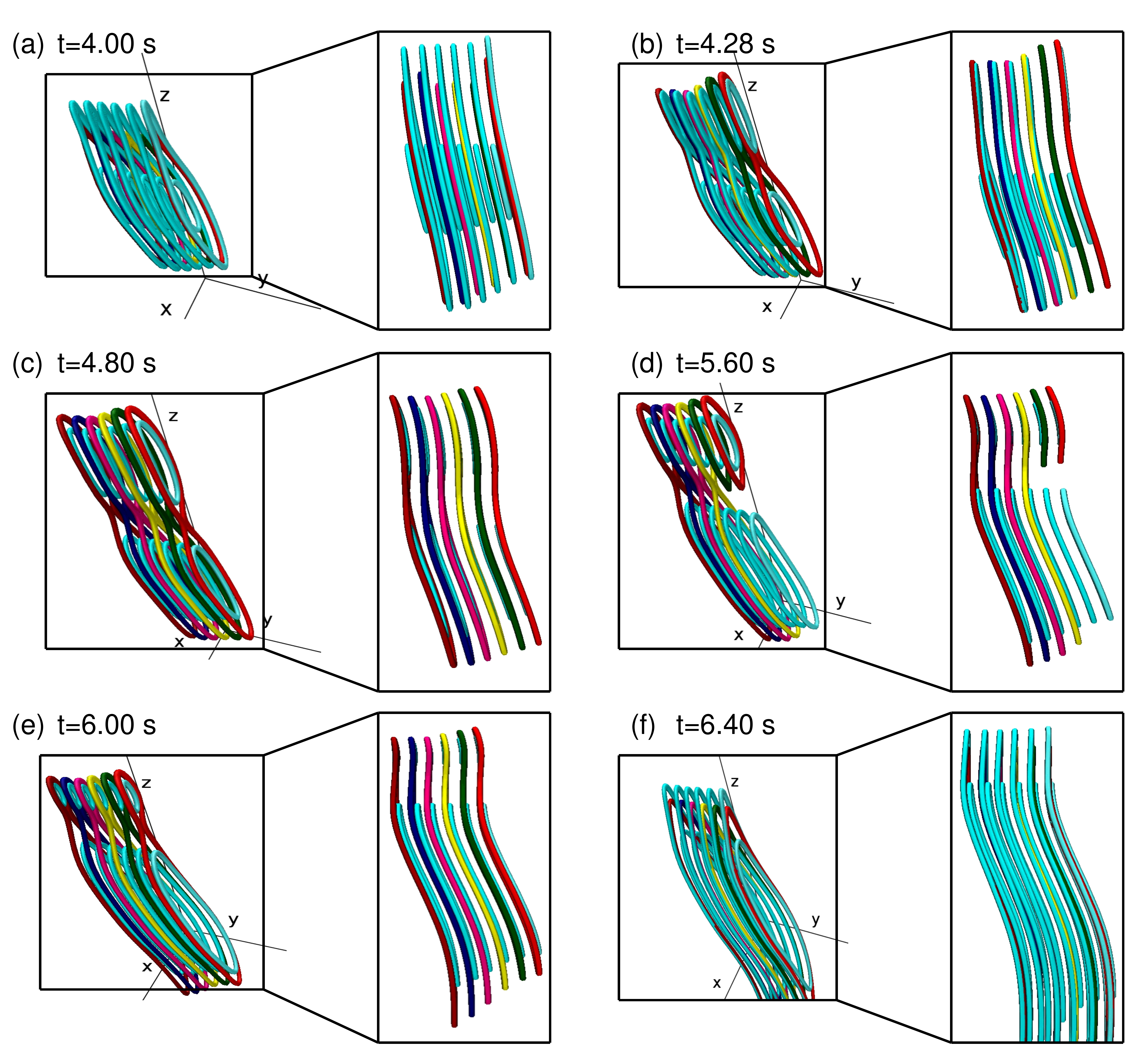}
\caption{Panels (a) to (f) show the MFL evolution for $d_0=2$ (HMHD) case, from two different angles to highlight the generation of two MFTs constituted by disjoint MFLs.  The islands look like ``figure 8" structure; Panels (b) to (d). The side view of the MFLs are shown in the insets, highlighting their undulated geometry. The three black lines in the 
background represent the three axes.
}\label{chips}
\end{figure*}

\begin{figure*}[h]
\includegraphics[width=1\textwidth]{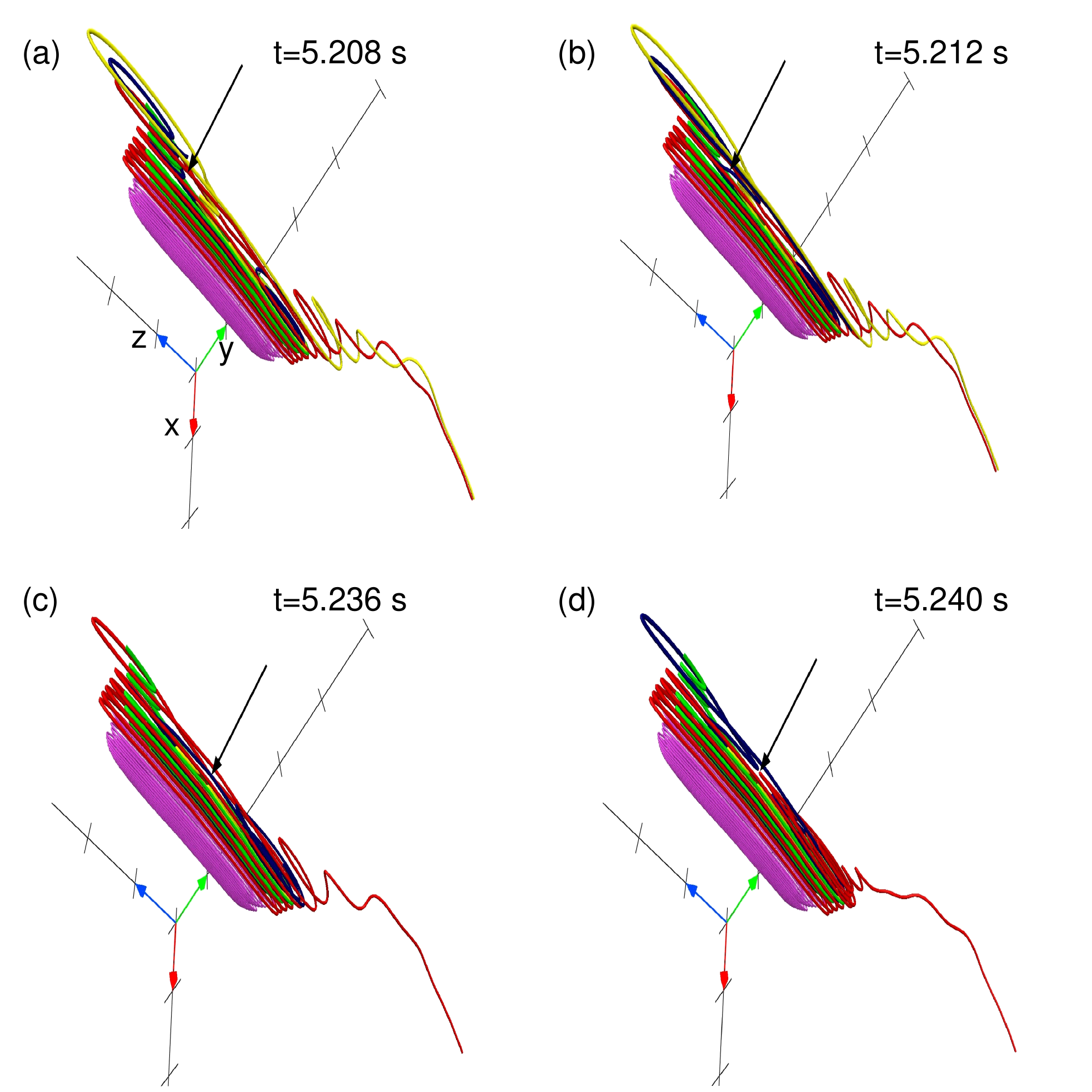}
\caption{Panels (a) and (b) show the topology of the MFLs for the $d_0=2$ (HMHD) evolution, prior and after the internal reconnection of the dark blue colored MFL (marked by the arrow). Panels (c) and (d) depict the topology of MFLs prior and after the internal reconnection of blue and red color MFLs within rope, marked by arrow. (An animation of this figure is available.)}\label{ropewhole}
\end{figure*}

\begin{figure*}[h]
\includegraphics[width=1\textwidth, height=0.8\textwidth]{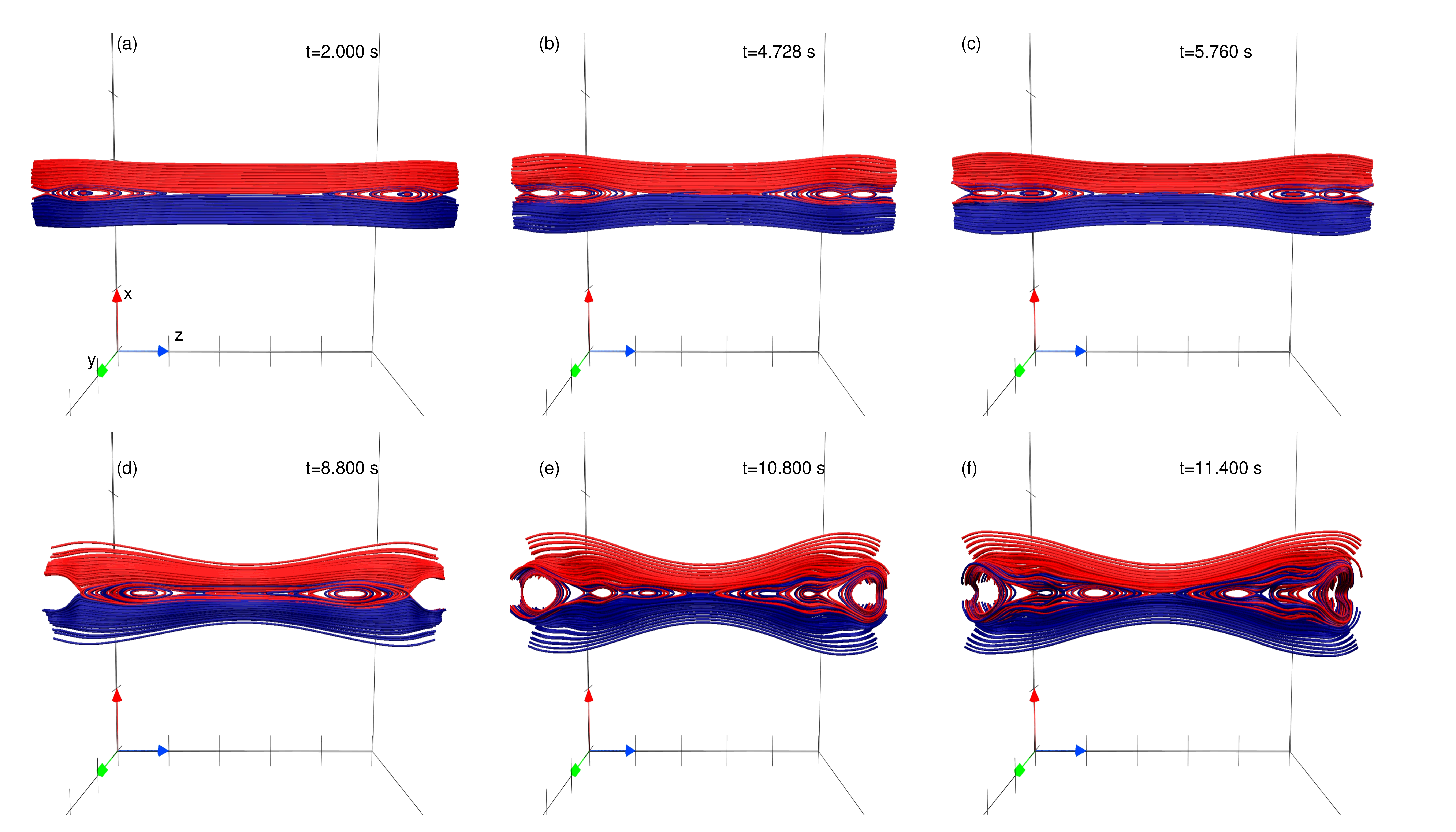}
\caption{MFL evolution for $d_0=2$ (HMHD), projected on $y$ constant plane. Panel (a) 
depicts development of two primary
magnetic islands. Panels (b) and (c) show their further breakage into secondary islands. Panels (d) to (f) 
show generation of an  X type neutral point by subsequent merging of the two islands. (An animation of this figure is available.)
}\label{2Dsind02}
\end{figure*}

For better clarity the MFLs evolution is further detailed in Figures \ref{chips} and \ref{ropewhole}.  In Figure \ref{chips} important is the development of two MFTs constituted by disjointly stacked magnetic islands. The islands are undulated and appear much earlier  
compared to the $d_0=0$ case, indicating the faster reconnection. Notable is also the creation of flux ropes 
where  a single helical MFL makes 
a large number of turns as the out-of-plane field $B_y$ develops (Figure \ref{ropewhole}). In principle, the 
MFL may ergodically span the MFS, if the ``safety factor" $q=r B_y/{\mathcal L} B_T$ is not a 
rational number \citep{Freidberg}; here $r$ and ${\mathcal L}$ are the radius and length of the 
rope, respectively, and $B_T =\sqrt{B_x^2+B_y^2}$ .
 Further evolution breaks the flux rope into secondary ropes by internal MRs---i.e.,
reconnections between
MFLs constituting the rope---shown in panels (a) to (d) of Figure \ref{ropewhole}, where two oppositely directed sections of the given MFLs reconnect (location marked by arrows in the Figure \ref{ropewhole}). Since most of the contemporary Hall simulations are in 2D, 
in Figure \ref{2Dsind02} we plot the projection of MFLs depicted in Figure \ref{twovp} on 
$y=0.5$ plane. The corresponding evolution is visibly similar to the generation of secondary
islands \citep{chenshi}, and their later coalescence as envisioned by \citet{Shibata}.

To complete the benchmarking, we repeated the numerical experiment described in \citet{Hubabook},
where wave propagation in the presence of the Hall forcing is explored. Notably, the EULAG-MHD being incompressible, 
we only concentrate on the whistler wave. The experimental setup is identical to that in the \citet{Hubabook}; the
ambient field is given by 

\begin{equation}
{\bf{B}}=B_0 \cdot \hat{e_z}~,
\label{ambB}
\end{equation}
whereas the perturbations are
\begin{eqnarray}
& &\delta B_x=\delta B_0 \sin \left(\frac{2\pi mz}{L_0}\right)~, \\
& &\delta B_y=\delta B_0 \cos \left(\frac{2\pi mz}{L_0}\right)~,
\label{pertB}
\end{eqnarray}

\noindent where $B_0=1000$, $\delta B_0=10$ and $L_0=7\pi$. The mode number is represented by m. The simulations are carried out on a computational domain of size $128\times 128\times 128$
and the dimensionless HMHD equations (discussed shortly) are employed. The analytical and the numerical 
frequencies obtained for various modes are listed in the Table-2, confirming the simulations to replicate
the analytical calculations fairly well.

\begin{deluxetable*}{ccc}[h]
\tablenum{2}
\tablecaption{List of parameters for the wave simulation\label{tab:tabi}}
\tablewidth{0pt}
\tablehead{
\colhead{m} & \colhead{Analytical frequency ($\omega_A$)} & \colhead{Numerical frequency ($\omega_N$)}  
}
\startdata
2 & 26.12 & 27.50 \\
\bigskip
3 & 132.237 & 141.88 \\
\bigskip
4 & 417.92 & 430.04 
\enddata
\end{deluxetable*}

\subsection{Activation of magnetic flux rope from bipolar magnetic field}
Sheared bipolar magnetic field is ubiquitous in solar plasma and plays an important 
role in the onset of  solar transients. In brief, the coronal mass ejection models require 
magnetic flux ropes (MFRs) to confine plasma. Destabilized from its equilibrium, 
as the MFR ascends with height---it stretches the overlaying MFLs. The ascend of the
rope decreases the magnetic pressure below it which, in turn sucks in 
more MFLs. These non-parallel MFLs reconnect and the generated outflow further pushes the 
MFR up. Details about the coronal mass ejection can be found 
in the review article by \citep{Chen2011}. Recently, \citet{Sanjay2016} numerically simulated the above 
scenario to explain generation and dynamics of a MFR beginning from  initial sets of 
sheared and twisted MFLs.  In the following, we conduct simulations to numerically explore
such activations of MFRs in the presence of the Hall term. For this purpose, 
we employ dimensionless form of the HMHD equations achieved through the following normalizations
\citep{Sanjay2016}

 \begin{equation}
 \label{norm}
 {\bf{B}}\longrightarrow \frac{{\bf{B}}}{B_0},\quad{\bf{v}}\longrightarrow \frac{\bf{v}}{v_A},\quad
  L \longrightarrow \frac{L}{L_0},\quad t \longrightarrow \frac{t}{\tau_a},\quad
  p  \longrightarrow \frac{p}{\rho {v_a}^2}~. 
 \end{equation}
The constant $B_0$ is kept arbitrary, whereas $L_0$ is fixed to the system size. Further, $v_A 
\equiv B_0/\sqrt{4\pi\rho_0}$ is the Alfv\'{e}n speed, where $\rho_0$ is a constant mass density. The  $\tau_A
$ and $\tau_\nu$ are respectively the Alfv\'{e}n transit time ($\tau_A=L_0/v_A$) and viscous diffusion time 
scale ($\tau_\nu= L_0^2/\nu$). The kinematic viscosity is denoted by $\nu$.
The normalized equations can be readily visualized by setting $1/(4\pi\rho_0)=1$ and 
$\mu_0=\tau_A/\tau_\nu$ in (\ref{momtransf}), $d_0/(4\pi)=\delta_i/L$ in (\ref{induc}),
while keeping (\ref{incompv}) and (\ref{incompb}) unchanged. The ratio  $\tau_A/\tau_\nu$ is an 
effective viscosity of the system which,  along with the other forces, 
 influences the magnetofluid evolution.

\begin{figure*}[h]
\includegraphics[width=1\textwidth]{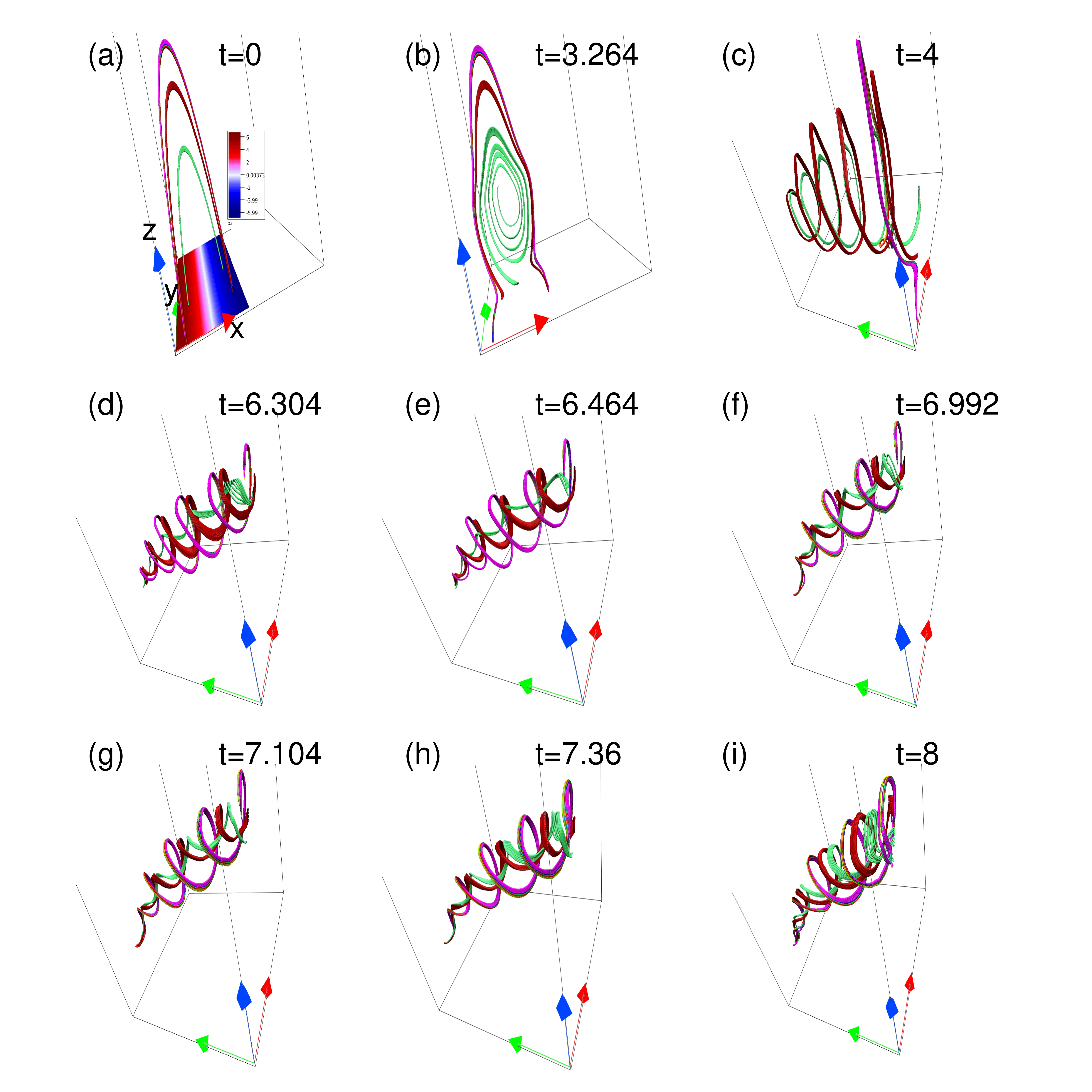}
\caption{Panel (a) shows the initial bipolar sheared arcade configuration along with polarity inversion line, Panels (b) and (c) show the formation of magnetic flux rope. Panels (d) to (i) represent the further evolution of the magnetic flux rope with a tilted axis (along $y$) of it for $\delta_i/L=0$ (MHD) case.}\label{FRd00}
\end{figure*}

 The simulation is initiated with the magnetic field
given in  \citet{Sanjay2016}
\begin{eqnarray}
\label{inirope}
B_x & & = k_z \sin (k_x x) \exp \left(\frac{-k_z z}{s_0}\right)~,\\
B_y & & = \sqrt{k_x^2-k_z^2} \sin (k_x x) \exp \left(\frac{-k_z z}{s_0}\right)~,\\
B_z & & = s_0 k_x  \cos (k_x x) \exp \left(\frac{-k_z z}{s_0}\right)~,
\end{eqnarray}
with $k_x=1.0$, $k_z=0.9$ and $s_0=6$. The  effective viscosity and mass density are set to $\tau_A/\tau_\nu=2\times 10^{-5}$ and
$\rho_0=1$, respectively. The MFLs are depicted in panel (a) of Figure \ref{FRd00} which are sheared bipolar loops having a 
straight Polarity Inversion Line (PIL) and no field-line twist. 
For simulations, a physical domain
of  the extent [$\{0,2\pi\},\{0,2\pi\},\{0,8\pi\}$] is resolved on the computational domain
of size $64\times 64\times 128$, making the spatial step sizes 
$\delta x=\delta y=0.0997,~\delta z=0.1979$. The temporal step size is 
$\delta t=16\times 10^{-4}$. The initial state is assumed to be motionless
and open boundary conditions are employed. 
The simulations are carried out for $\delta_i/L_0=0$ and $\delta_i/L_0=0.04$,
having a simulated physical time of $7000\tau_A\delta t $. The arbitrary $B_0$
can be selected such that the Alfv\'{e}n transit time, $\tau_A \in\{1,10\}$ s 
makes the simulated time, $11.2$ s to $112$ s consistent with the beginning of the
impulsive phase of a flare $100$ s to $1000$ s.
The simulation parameters for MFR are listed in the Table 3.
 \begin{deluxetable*}{ccccccc}[h]
\tablenum{3}
\tablecaption{List of parameters for simulation with bipolar sheared magnetic arcade initial field\label{tab:tabi}}
\tablewidth{0pt}
\tablehead{
\colhead{$\rho_0$}  & \colhead{$\delta_i$} & \colhead{L} 
& \colhead{$\frac{\delta_i}{L}$} & \colhead{Simulation Box Size} & \colhead{Resolution} & \colhead{Effective Viscosity $\frac{\tau_A}{\tau_{\nu}}$}
}
\startdata
1.0 & 1.005 & $8\pi$ & 0.04 & $2\pi\times 2\pi\times 8\pi$ & $64\times 64\times 128$ & 2$\times$10$^{-5}$
\enddata
\end{deluxetable*}
 The evolution onsets as the Lorentz force 
\begin{eqnarray}
(\textbf{J}\times\textbf{B})_x & & = \left[-k_x(k_x^2-k_z^2)+k_x s_0 \left(s_0 k_x^2-\frac{k_z^2}{s_0}\right)\right]\times \sin^2 (k_x x)\exp\left(-\frac{2k_z z}{s_0}\right)~,\\
(\textbf{J}\times\textbf{B})_z & & = \left[\frac{k_z}{s_0}(k_x^2-k_z^2)-k_z  \left(s_0 k_x^2-\frac{k_z^2}{s_0}\right)\right]\times \frac{\sin (2k_x x)}{2}\exp\left(-\frac{2k_z z}{s_0}\right)~, 
\end{eqnarray}
pushes oppositely directed segments of MFLs toward each other, generating the neck at
$t=3.264$, panel (b) of the Figure \ref{FRd00}---demonstrating the MFL dynamics for $d_0=0$. The MRs at the 
neck generate the MFR---which we refer as the primary MFR (panel (c) of the Figure \ref{FRd00}). Further 
evolution preserves the primary MFR by not allowing it to go through any internal MRs. 
Notably, the rope loses its initial symmetry along the $y$ direction by a marginal amount which,
 we attribute to the open boundary conditions. Nevertheless, the rope rises uniformly about
a slightly inclined axis.

\begin{figure*}[h]
\includegraphics[width=1\textwidth]{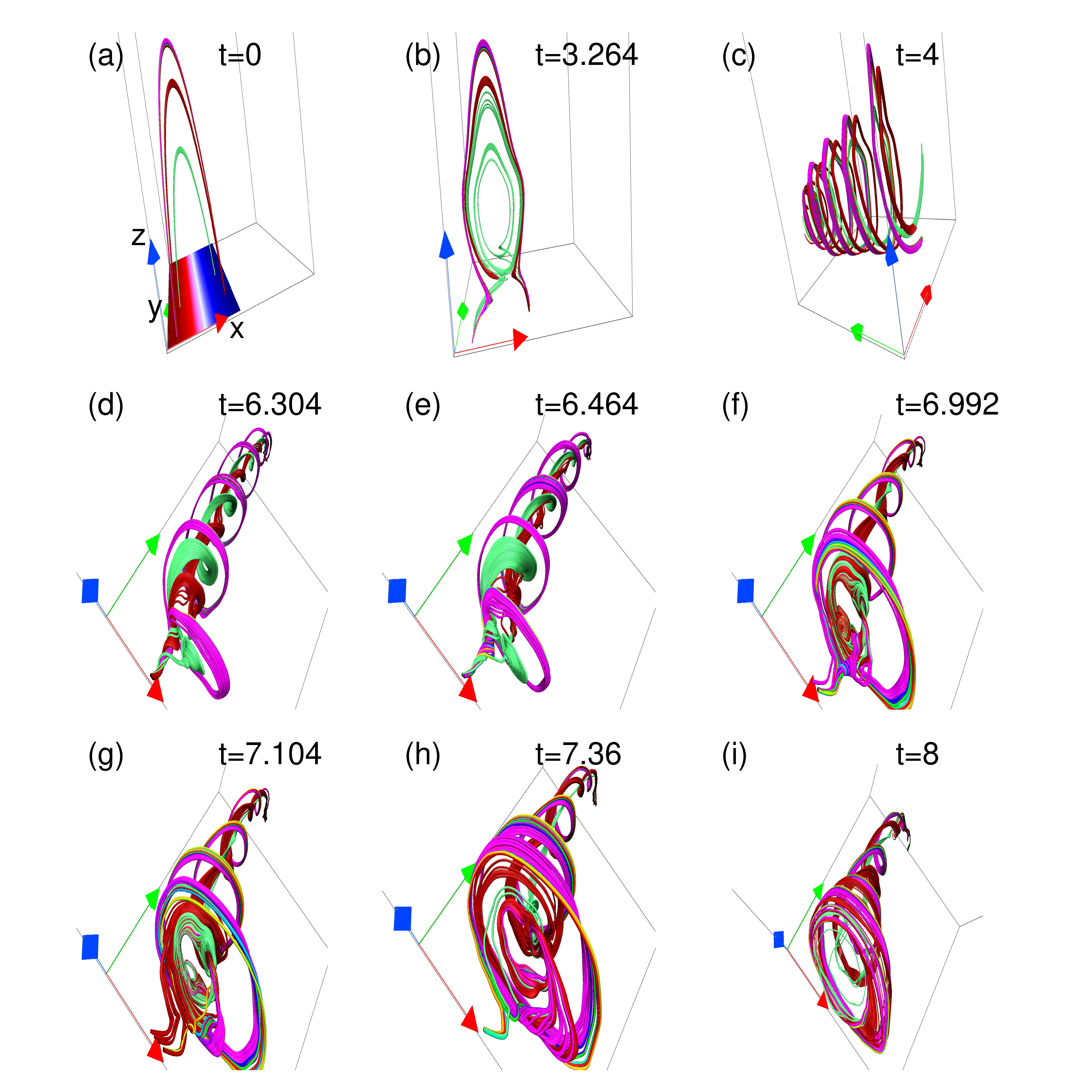}
\caption{Panels (a) to (i) show the topology of MFLs in their evolution for $\delta_i/L=0.04$ under the Hall forcing.
Important is the similarity of the dynamics leading to the formation of primary MFR which generates 
at a similar instant as the primary MFR in the absence of the Hall forcing. (A combined animation of Figure \ref{FRd00} and Figure \ref{d0p04rope} is available.) }\label{d0p04rope}
\end{figure*}

\begin{figure*}[h]
\includegraphics[width=1\textwidth]{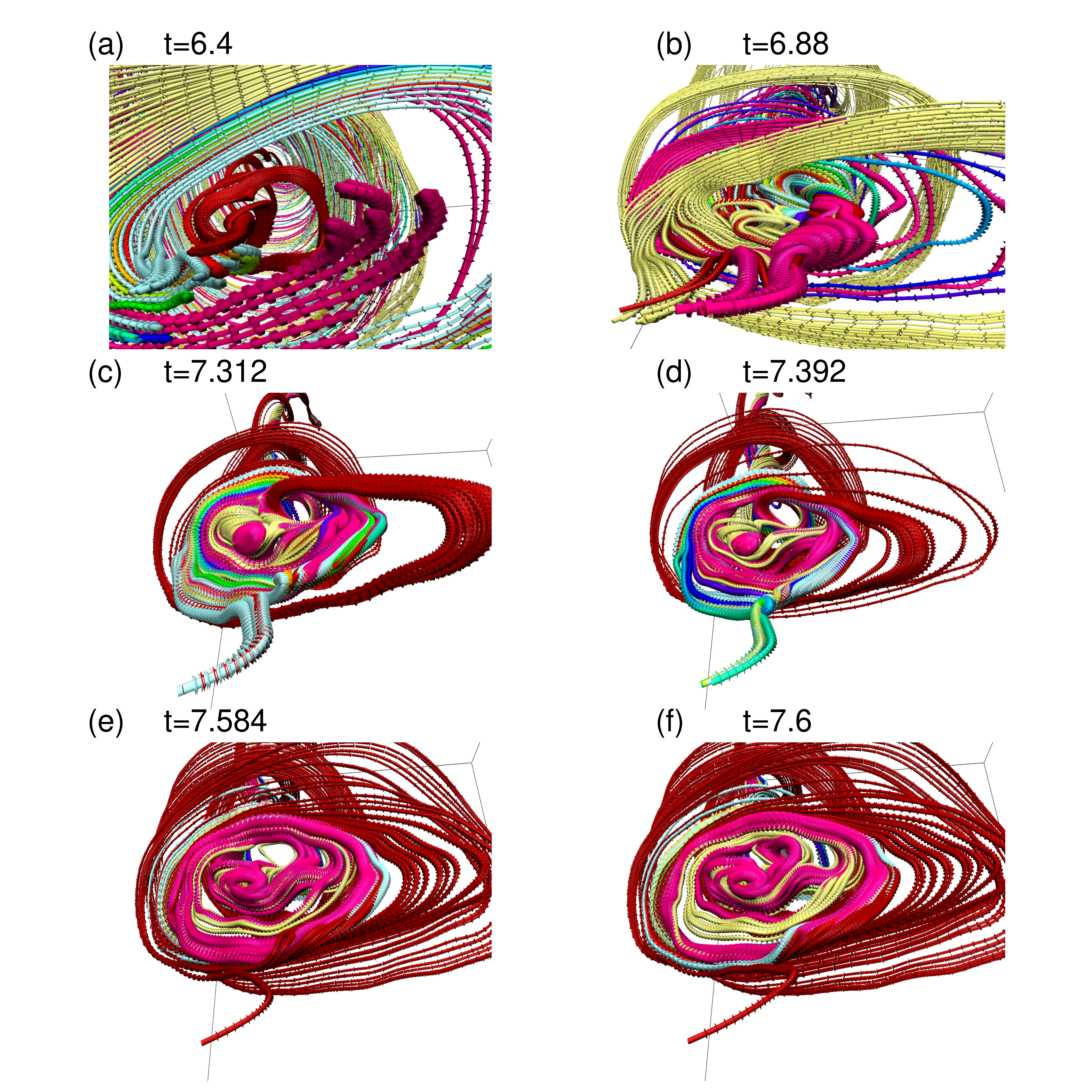}
\caption{Sequence of MFL evolution under the Hall forcing ($\delta_i/L=0.04$ case), zoomed to reveal
intricate magnetic topologies generated by the MRs. Formation of the ``figure 8" kind magnetic structures
(panels (a) to panel (f))---the magnetic islands---can be seen clearly. Importantly, such intricate topologies
are absent in the MFR evolution without the Hall forcing. (An animation of the evolution from $t=7.312$ onwards is available.)
}\label{ropeIR}
\end{figure*}

The MFL evolution for $\delta_i/L=0.04$ is exhibited in Figure \ref{d0p04rope}. The selected value 
is on the order of the coronal value quoted in Introduction and optimizes the computation. The primary MFR 
develops at $t=4$,  which is similar to the instant at which the primary MFR was generated for the $\delta_i/
L=0$ case. The overall dynamics leading to the primary MFR 
also remains similar to the one without the Hall forcing. The similar dynamics and the near-simultaneity in 
the onset of the the primary MFR in both cases indicate the large scale dynamics, 
i.e., the dynamics before or away from MRs, to be insensitive to the particular Hall forcing.
However, there are conspicuous differences between the MHD and HMHD realizations of the MFR morphology. In the HMHD case the primary MFR undergoes 
multiple internal MRs highlighted in  Figure 
\ref{ropeIR},
leading to MFL morphologies which when projected favorably look like magnetic islands similar to those 
found in the sinusoidal simulation. A swirling motion is also observed;  cf. panels 
(a) to (f) of Figure \ref{ropeIR} (better visualized in the animation). Noteworthy, swirling motion during evolution of 
a prominence eruption has been observed \citep{Pant_2018}.

\begin{figure*}[h]
\includegraphics[width=1\textwidth]{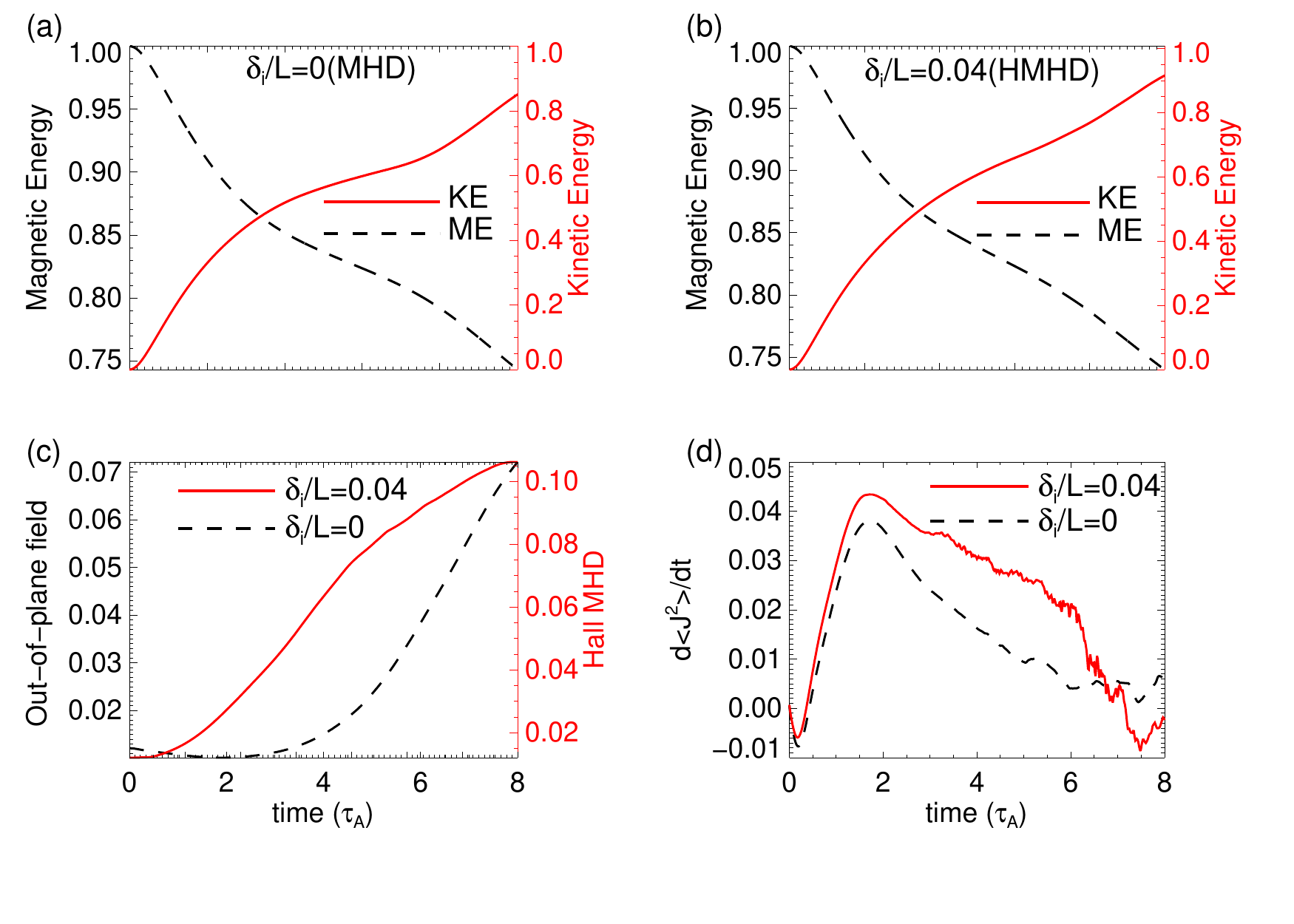}
\caption{Panels (a) and (b) show the evolution of normalized (with the initial total energy) grid averaged magnetic energy (black dashed curve) and kinetic energy (red solid curve) for $\delta_i/L=0$ (MHD) and $\delta_i/L=0.04$ (HMHD) respectively. Panel (c) shows the evolution of grid averaged out-of-plane magnetic field for $\delta_i/L=0$ (MHD) with black dashed curve and $\delta_i/L=0.04$ (HMHD) with red solid curve respectively. Also in Panels (a) to (c), the scales for the solid and the dashed curves are spaced at right and left respectively.  Panel (d) represents grid averaged rate of change of total current density for $\delta_i/L=0$ (black dashed curve) and $\delta_i/L=0.04$ (red solid curve) respectively. Important are the generation of the out-of-plane magnetic field along with
small but abrupt changes in time derivative of the total volume current density in HMHD simulation.}\label{FRoverall}
\end{figure*}

To complete the analyses, we plot the overall evolution of 
magnetic and kinetic energies, amplitude of the out-of-plane field and the rate of change
of the total volume current density in panels (c) and (d) of the Figure \ref{FRoverall}. 
The similarity of the energy curves in the presence and absence of the Hall forcing is a reminiscent 
of the fact that the Hall term does not affect the system energetics directly.
Importantly, the out-of-plane magnetic field (approximated by the axial magnetic field $B_y$) is larger than that in the absence of the 
Hall forcing, in accordance with the expectation. Further, contrary to its smooth variation in the 
MHD case, the rate of change of total volume current density in HMHD goes through small but abrupt changes. 
Such abrupt changes may correspond to a greater degree of impulsiveness \citep{BhattacharjeeReview}.

\par To check the dependence of the above findings on the grid resolution, we have carried out 
auxiliary simulations with $32\times32\times64$ grid resolution, spanning the same physical domain
with all the other parameters kept identical (not shown). The findings are similar to those at the higher resolution. In particular, they evince the nearly simultaneous formation of the primary MFR, with and without the Hall forcing, 
through the similar dynamical evolution. Also, breakage of the primary MFR through internal MRs is found in 
presence of Hall forcing
whereas no such breakage is seen in the absence of the Hall forcing. The identical dynamics in two separate 
resolutions indicate the findings to be independent of the particular resolution used.

\section{Summary}
\label{conclusion}
The Eulerian-Lagrangian model EULAG-MHD has been extended to the HMHD regime, by modifying the 
induction equation to include the Hall term. Subsequently, benchmarking is done with an initially sinusoidal 
magnetic field, symmetric in the $y$-direction of the employed Cartesian coordinate system.  The choice of the 
field is based on its simplicity and non-force-free property to exert Lorentz force on the magnetofluid at $t=0$. 
Moreover, the selected field provides an opportunity to independently verify the physics of HMHD without 
repeating the more traditional computations related to the Harris equilibrium or the GEM challenge.
Simulations are carried out in the absence and presence of the Hall term. 
In the absence of the Hall term the magnetic field maintains its symmetry as MRs
generate magnetic flux tubes made by disjoint MFLs. With the Hall term, the  
evolution becomes asymmetric
and 3D due to the development  of magnetic field which is 
directed out of the reconnection plane. This is in concurrence
with  earlier simulations. Along with the flux tube, MRs also generate magnetic flux rope in the HMHD. 
 When viewed along the $-y$ direction, the rope and the tube 
appear as magnetic islands. Further evolution, leads to breakage of the primary islands 
into secondary islands and later, their coalescence. The results, overall, agree with the 
existing scenarios of Hall-reconnection based on physical arguments and other recent simulations
including those on the GEM challenge.  An important finding is the formation of complex 3D
magnetic structures which can not  
be apprehended from 2D models or calculations although  their projections agree with the latter.
Alongside, we have numerically explored the Whistler mode propagation vis-a-vis its analytical model
and found the two to be matching reasonably well.

We have further carried out simulations where the EULAG-MHD is used to simulate the onset and dynamics
of a MFR initiating from a sheared magnetic arcade. Such computations are 
relevant in understanding the solar eruptions. 
Simulations conducted with and without the 
Hall term are compared once more. Once again a reasonable maintenance of symmetry is  observed in the 
standard MHD 
simulation, whereas a clear symmetry-breaking---leading to generation of 3D magnetic structures---appears to 
be a signature of the Hall effect. In HMHD the MFR evolves through a series of complex geometries
while rotating along its axis. 
When viewed favorably, it appears to contain  structures like the ``figure 8", which is the 
result of internal reconnection within MFR. Notably, the magnetic and kinetic energies, in the presence and absence of the Hall forcing, behave almost identically---consistent with the theoretical 
understanding that the  Hall term does not directly change 
the magnetic energy. Moreover, we have performed and analyzed the simulations with half the resolution and found their results to be similar to the reference results. The same process of the primary MFR formation and its further breakage through multiple internal magnetic reconnections does confirm the independency of the results on the grid resolution.

Overall, the extended EULAG-MHD is giving results in accordance with the theoretical expectations 
and other contemporary simulations. The Hall simulation documenting activation of a magnetic flux rope 
from initial sheared arcade field lines is of particular importance and is a new entry to 
the ongoing research. It shows that Hall magnetohydrodynamics can account for a richer complexity 
during evolution of the rope by breaking any pre-existing 2D symmetry, thus opening another degree 
of freedom for the MFL evolution. The resulting local breakage of the rope is intriguing 
by itself and calls for further research.

\begin{acknowledgments}

The authors thank an anonymous referee for providing insightful comments and 
suggestions which increased the scientific content and readability of the paper. The simulations are performed using the 100TF cluster Vikram-100 at Physical Research Laboratory, India. We wish to acknowledge the
visualization software VAPOR (www.vapor.ucar.edu), for generating
relevant graphics. NCAR is sponsored by the National Science Foundation.

\end{acknowledgments}


\begin{thebibliography}{}

\bibitem[\protect\citeauthoryear{{Alfv{\'e}n}}{{Alfv{\'e}n}}{1942}]{Alfven}
{Alfv{\'e}n}, H. 1942, \nat, 150, 405

\bibitem[\protect\citeauthoryear{{Aschwanden}}{{Aschwanden}}{2005}]{Aschwanden}
{Aschwanden}, M.~J. 2005, {Physics of the Solar Corona. An Introduction with
  Problems and Solutions (2nd edition)}

\bibitem[\protect\citeauthoryear{{Bhattacharjee}}{{Bhattacharjee}}{2004}]{BhattacharjeeReview}
{Bhattacharjee}, A. 2004, \araa, 42, 365

\bibitem[\protect\citeauthoryear{{Bhattacharyya}, {Low}, \&
  {Smolarkiewicz}}{{Bhattacharyya} et~al.}{2010}]{RBCLOW}
{Bhattacharyya}, R., {Low}, B.~C.,  \& {Smolarkiewicz}, P.~K. 2010, Physics of
  Plasmas, 17, 112901

\bibitem[\protect\citeauthoryear{{Birn} et~al.}{{Birn} et~al.}{2001}]{BIRN}
{Birn}, J., et~al. 2001, \jgr, 106, 3715

\bibitem[\protect\citeauthoryear{{Charbonneau} \&
  {Smolarkiewicz}}{{Charbonneau} \& {Smolarkiewicz}}{2013}]{Piotrscience}
{Charbonneau}, P.,  \& {Smolarkiewicz}, P.~K. 2013, Science, 340, 42

\bibitem[\protect\citeauthoryear{{Chen}}{{Chen}}{2011}]{Chen2011}
{Chen}, P.~F. 2011, Living Reviews in Solar Physics, 8, 1

\bibitem[\protect\citeauthoryear{{Choudhuri}}{{Choudhuri}}{1998}]{Arnab}
{Choudhuri}, A.~R. 1998, {The physics of fluids and plasmas : an introduction
  for astrophysicists }

\bibitem[\protect\citeauthoryear{{Freidberg}}{{Freidberg}}{1982}]{Freidberg}
{Freidberg}, J.~P. 1982, Reviews of Modern Physics, 54, 801

\bibitem[\protect\citeauthoryear{Grinstein, Margolin, \& Rider}{Grinstein
  et~al.}{2007}]{Grinstein2007}
Grinstein, F., Margolin, L.,  \& Rider, W. 2007, Implicit Large Eddy
  Simulation: Computing Turbulent Fluid Dynamics, Cambridge University Press

\bibitem[\protect\citeauthoryear{Huba}{Huba}{2003}]{Hubabook}
Huba, J.~D. 2003, Hall Magnetohydrodynamics - A Tutorial, ed. J.~B{\"u}chner,
  M.~Scholer, \& C.~T. Dum (Berlin, Heidelberg: Springer Berlin Heidelberg),
  166

\bibitem[\protect\citeauthoryear{{Huba} \& {Rudakov}}{{Huba} \&
  {Rudakov}}{2002}]{Huba}
{Huba}, J.~D.,  \& {Rudakov}, L.~I. 2002, Physics of Plasmas, 9, 4435

\bibitem[\protect\citeauthoryear{{Kumar} \& {Bhattacharyya}}{{Kumar} \&
  {Bhattacharyya}}{2011}]{DKRB}
{Kumar}, D.,  \& {Bhattacharyya}, R. 2011, Physics of Plasmas, 18, 084506

\bibitem[\protect\citeauthoryear{{Kumar} et~al.}{{Kumar} et~al.}{2017}]{SK2017}
{Kumar}, S., {Bhattacharyya}, R., {Dasgupta}, B.,  \& {Janaki}, M.~S. 2017,
  Physics of Plasmas, 24, 082902

\bibitem[\protect\citeauthoryear{{Kumar} et~al.}{{Kumar}
  et~al.}{2016}]{Sanjay2016}
{Kumar}, S., {Bhattacharyya}, R., {Joshi}, B.,  \& {Smolarkiewicz}, P.~K. 2016,
  \apj, 830, 80

\bibitem[\protect\citeauthoryear{{Kumar}, {Bhattacharyya}, \&
  {Smolarkiewicz}}{{Kumar} et~al.}{2015}]{SKRB}
{Kumar}, S., {Bhattacharyya}, R.,  \& {Smolarkiewicz}, P.~K. 2015, Physics of
  Plasmas, 22, 082903

\bibitem[\protect\citeauthoryear{{Ma} \& {Bhattacharjee}}{{Ma} \&
  {Bhattacharjee}}{2001}]{MaBhattacharjee}
{Ma}, Z.~W.,  \& {Bhattacharjee}, A. 2001, \jgr, 106, 3773

\bibitem[\protect\citeauthoryear{{Mozer}, {Bale}, \& {Phan}}{{Mozer}
  et~al.}{2002}]{Mozer2002}
{Mozer}, F.~S., {Bale}, S.~D.,  \& {Phan}, T.~D. 2002, \prl, 89, 015002

\bibitem[\protect\citeauthoryear{{Nayak} et~al.}{{Nayak} et~al.}{2019}]{Ss2019}
{Nayak}, S.~S., {Bhattacharyya}, R., {Prasad}, A., {Hu}, Q., {Kumar}, S.,  \&
  {Joshi}, B. 2019, \apj, 875, 10

\bibitem[\protect\citeauthoryear{{Nayak} et~al.}{{Nayak} et~al.}{2020}]{Ss2020}
{Nayak}, S.~S., {Bhattacharyya}, R., {Smolarkiewicz}, P.~K., {Kumar}, S.,  \&
  {Prasad}, A. 2020, \apj, 892, 44

\bibitem[\protect\citeauthoryear{Pant et~al.}{Pant et~al.}{2018}]{Pant_2018}
Pant, V., Datta, A., Banerjee, D., Chandrashekhar, K.,  \& Ray, S. 2018, The
  Astrophysical Journal, 860, 80

\bibitem[\protect\citeauthoryear{{Parker}}{{Parker}}{1994}]{ParkerECS}
{Parker}, E.~N. 1994, Spontaneous current sheets in magnetic fields : with
  applications to stellar x-rays. International Series in Astronomy and
  Astrophysics, 1

\bibitem[\protect\citeauthoryear{{Prasad} et~al.}{{Prasad}
  et~al.}{2018}]{avijeet2018}
{Prasad}, A., {Bhattacharyya}, R., {Hu}, Q., {Kumar}, S.,  \& {Nayak}, S.~S.
  2018, \apj, 860, 96

\bibitem[\protect\citeauthoryear{{Prasad}, {Bhattacharyya}, \&
  {Kumar}}{{Prasad} et~al.}{2017}]{avijeet2017}
{Prasad}, A., {Bhattacharyya}, R.,  \& {Kumar}, S. 2017, \apj, 840, 37

\bibitem[\protect\citeauthoryear{{Priest}}{{Priest}}{2014}]{MHDpriest}
{Priest}, E. 2014, {Magnetohydrodynamics of the Sun}

\bibitem[\protect\citeauthoryear{{Priest} \& {Forbes}}{{Priest} \&
  {Forbes}}{2000}]{PF200}
{Priest}, E.,  \& {Forbes}, T. 2000, {Magnetic Reconnection}

\bibitem[\protect\citeauthoryear{Prusa, Smolarkiewicz, \& Wyszogrodzki}{Prusa
  et~al.}{2008}]{Prusa}
Prusa, J., Smolarkiewicz, P.,  \& Wyszogrodzki, A. 2008, Computers \& Fluids,
  37, 1193

\bibitem[\protect\citeauthoryear{{Shi} et~al.}{{Shi} et~al.}{2019}]{chenshi}
{Shi}, C., {Tenerani}, A., {Velli}, M.,  \& {Lu}, S. 2019, \apj, 883, 172

\bibitem[\protect\citeauthoryear{{Shibata} \& {Tanuma}}{{Shibata} \&
  {Tanuma}}{2001}]{Shibata}
{Shibata}, K.,  \& {Tanuma}, S. 2001, Earth, Planets, and Space, 53, 473

\bibitem[\protect\citeauthoryear{{Smolarkiewicz}}{{Smolarkiewicz}}{2006}]{Piotrsingle}
{Smolarkiewicz}, P.~K. 2006, International Journal for Numerical Methods in
  Fluids, 50, 1123

\bibitem[\protect\citeauthoryear{{Smolarkiewicz} \&
  {Charbonneau}}{{Smolarkiewicz} \& {Charbonneau}}{2013}]{PiotrJCP}
{Smolarkiewicz}, P.~K.,  \& {Charbonneau}, P. 2013, Journal of Computational
  Physics, 236, 608

\bibitem[\protect\citeauthoryear{{Sonnerup}}{{Sonnerup}}{1979}]{Sonnerup}
{Sonnerup}, B.~U.~{\"O}. 1979, {Magnetic field reconnection}, Vol.~3 45

\bibitem[\protect\citeauthoryear{{Westerberg} \&
  {{\r{A}}kerstedt}}{{Westerberg} \& {{\r{A}}kerstedt}}{2007}]{Westerberg}
{Westerberg}, L.~G.,  \& {{\r{A}}kerstedt}, H.~O. 2007, Physics of Plasmas, 14,
  102905

\end{thebibliography}
\end{document}